# The Problem of Distributed Consensus: A Survey

Stephen Wolfram*

*A survey is given of approaches to the problem of distributed consensus, focusing particularly on methods based on cellular automata and related systems. A variety of new results are given, as well as a history of the field and an extensive bibliography. Distributed consensus is of current relevance in a new generation of blockchain-related systems.*

*In preparation for a conference entitled "Distributed Consensus with Cellular Automata and Related Systems" that we're organizing with NKN (= "New Kind of Network") I decided to explore the problem of distributed consensus using methods from A New Kind of Science (yes, NKN "rhymes" with NKS) as well as from the Wolfram Physics Project.*

## A Simple Example

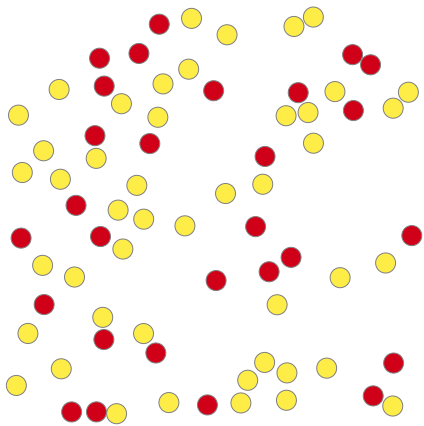

Consider a collection of "nodes", each one of two possible colors. We want to determine the majority or "consensus" color of the nodes, i.e. which color is the more common among the nodes.

---

One obvious method to find this "majority" color is just sequentially to visit each node, and tally up all the colors. But it's potentially much more efficient if we can use a distributed algorithm, where we're running computations in parallel across the various nodes.

One possible algorithm works as follows. First connect each node to some number of neighbors. For now, we'll just pick the neighbors according to the spatial layout of the nodes:

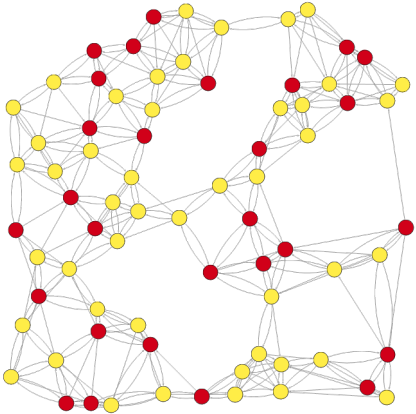

The algorithm works in a sequence of steps, at each step updating the color of each node to be whatever the "majority color" of its neighbors is. In the case shown, this procedure converges after a few steps to make all nodes have the "majority color" (which here is yellow)—or in effect "agree" on what the majority color is:

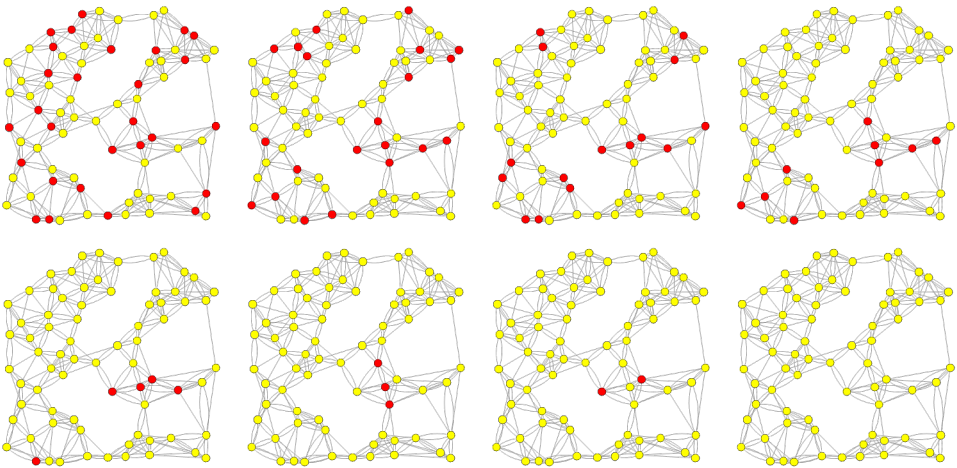

This is a simple example of a distributed consensus algorithm in action. The challenge we'll discuss here is to find the most efficient and robust such algorithms.

# The Background

In any decentralized system with computers, people, databases, measuring devices or anything else one can end up with different values or results at different "nodes". But for all sorts of reasons one often wants to agree on a single "consensus" value, that one can for example use to "make a decision and go on to the next step".

Blockchains are one example of systems that need this kind of consensus to "finish each block". Traditional blockchains achieve consensus through what amounts to a centralized mechanism (even though there are multiple "decentralized" copies of the blockchain that is produced).

But there are now starting to be distributed analogs of blockchains that need distributed consensus algorithms. And the main inspiration for the algorithms being developed are cellular automata (and to a lesser extent spin systems in statistical mechanics).

One issue is to make the algorithm as efficient as possible. Another is to make it as robust as possible, for example with respect to random noise—or malicious errors—introduced at or between nodes.

The amount of random noise can be thought of as something like a temperature. And at least in certain cases there can be a "phase transition" so that below a certain "temperature" there can be zero effect on the consensus output—implying robustness to a certain level of noise.

Some of what happens can be studied using methods from standard equilibrium statistical physics. But in most cases one has to take account of the time dependence or evolution of the system, leading to something like a probabilistic cellular automaton (closely related to directed percolation, dynamic spin systems, etc.).

As I'll discuss below, in the early days of computing, there was great interest in synthesizing reliable systems out of unreliable components. And by the 1960s there was study first of neural nets and then of cellular automata with probabilistic elements. And some surprising results were obtained that showed that cellular automata could be set up that would be robust with respect to a certain nonzero level of noise.

One feature of cellular automata is that their elements are all assumed to be arranged in a definite array, and to be updated in parallel "at the same time" in a sequence of steps. For many practical applications, however, one instead wants elements that are connected in some kind of graph (that may even be dynamic), and that are in general updated asynchronously, in no particular order.

The simple example we gave above is a graph cellular automaton: the connections between elements are defined by a graph, but the updates are all done synchronously at each step. In the past, it's been difficult to analyze the more general setup where there is no rigid notion of either space or time. But this is exactly the setup in our new Physics Project, and so there's now the potential to use its formalism and results (as well as intuition imported from physics) to make further progress.

## Deterministic Cellular Automata

To start getting some intuition for the problem of distributed consensus, let's consider the following very simple setup. We have a line of cells, each with one of two possible colors. Then we update the colors of these cells in a sequence of steps, based on a local rule which depends on neighboring cells. This system is a one-dimensional cellular automaton—of the kind that I started studying more than 40 years ago.

We imagine that the initial condition involves a fraction $p$ of red cells. We want all the cells to turn red if $p > \frac{1}{2}$, and all of them to turn yellow if $p < \frac{1}{2}$. The most obvious rule that might achieve this would just replace each cell by the majority color in its neighborhood (rule 232 in my numbering scheme):

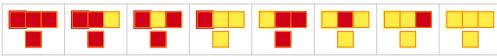

Here's what rule 232 does starting with 70% red cells in a "random configuration":

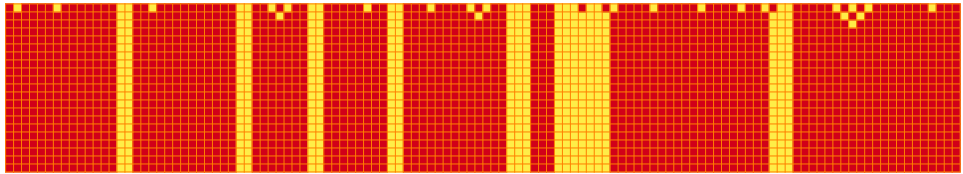

As we can see, it manages to achieve a little "local consensus", but ultimately it's not successful at reaching a "global consensus" in which all cells are the same color.

And we might imagine that there'd be no rule for a 1D deterministic cellular automaton that would lead to global consensus (or be able to solve the "density classification problem" of deciding whether the density of initial red cells is above or below 50%). But it turns out that this isn't true. And for example in 1978 the following "radius 3" rule (operating on size-7 neighborhoods) was constructed (the "GKL rule"):

{l3_, _, l1_, c_, r1_, _, r3_} :→ If[If[c == 0, r1 + r3, l1 + l3] + c ≥ 2, 1, 0]

Here's what this rule does with 60% red cells:

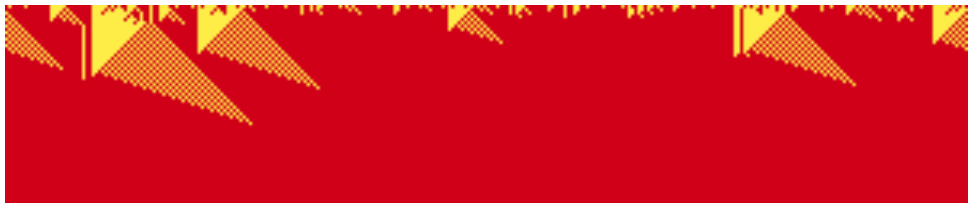

And here's what it does with 40% red cells:

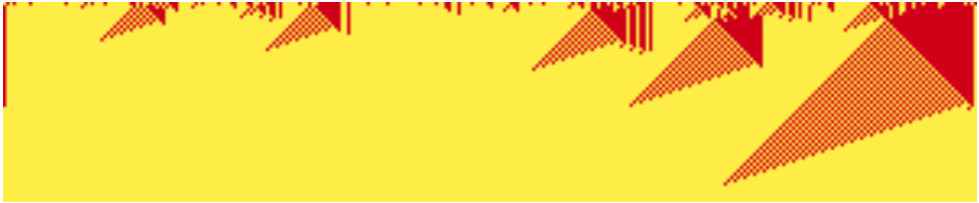

In both these cases, the rule successfully achieves "global consensus". And in fact one can prove that this rule will always do this, at least after sufficiently many steps. Here's a plot of how the density evolves as a function of time for different initial densities:

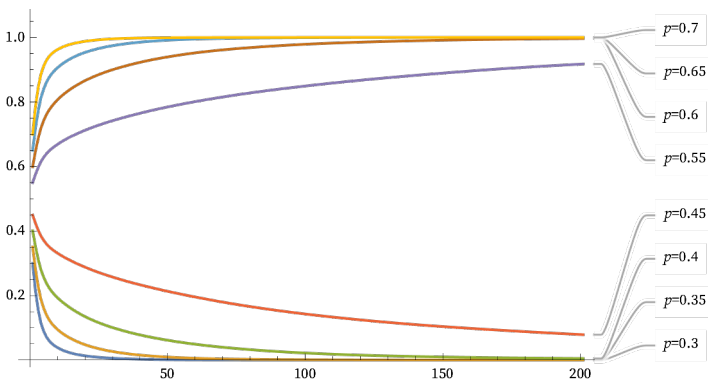

And what we see is that there's what looks like a phase transition: for initial density $p < 0.5$, the final density is exactly 0, while for initial density $p > 0.5$, it's instead exactly 1.

What happens precisely at $p = 0.5$? In a sense the cellular automaton "can't make up its mind" and on an infinite line it generates an infinite nested sequence of domains that alternate between 0 and 1:

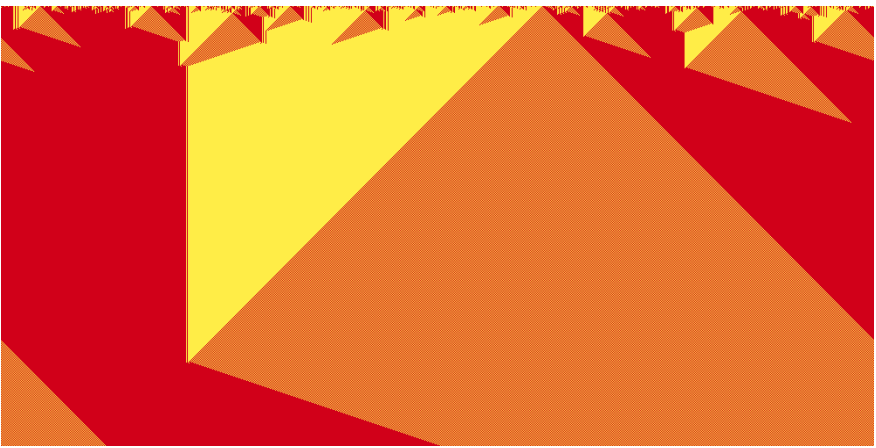

This nested structure is typical of what's seen in critical phenomena in statistical physics, and in fact cellular automata like this are the very simplest examples of "true" phase transitions that I know. (Like other phase transitions, these don't become "sharp" except in infinite systems. In typical statistical mechanics one doesn't get phase transitions in 1D, but that's a consequence of the assumption of microscopic reversibility, which doesn't apply to cellular automata like this.)

So what other cellular automaton rules achieve consensus like this? There are no radius-1 rules that work. And if one searches all $2^{32}$ radius-2 rules (as I did for *A New Kind of Science*), the best one finds are a handful of examples that achieve "approximate consensus" in the sense that most, though not all, of the cells go to the "majority value" (this is the $r = 2$ rule 4196304428, for $p = 0.6$):

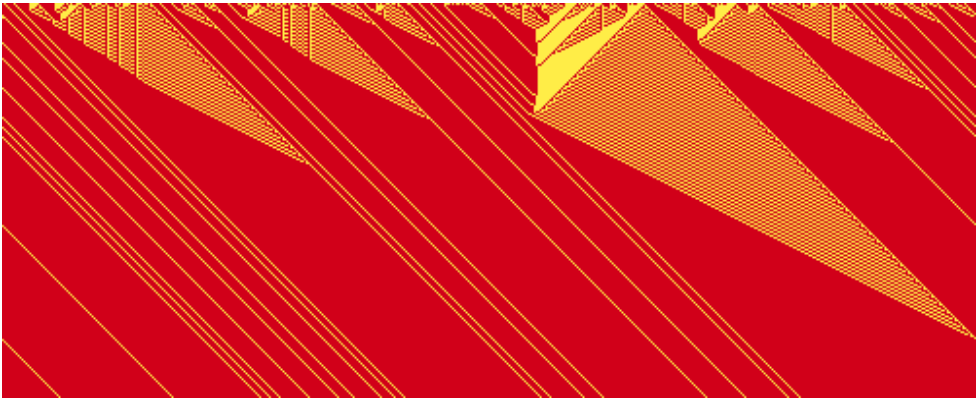

By the way, among radius-1 rules, there is rule 184 (often used as a basic model of road traffic flow), which doesn't achieve consensus on "overall density", but does do so with respect to left- and right-moving stripes, here with the nested pattern generated when $p = 0.5$:

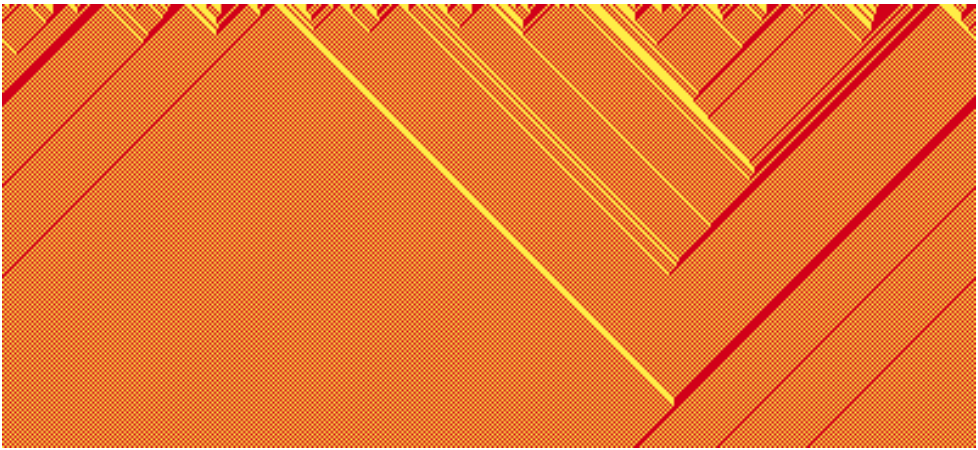

What about "achieving consensus faster"? Here's a comparison of our original GKL rule with another radius-3 rule (discovered by genetic programming methods) whose average consensus time is shorter:

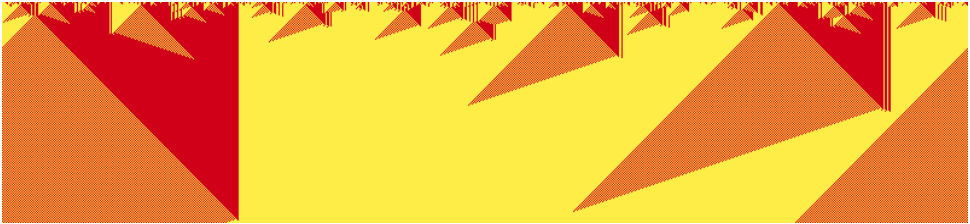

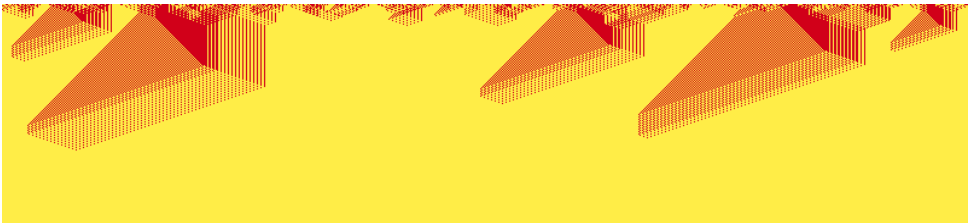

It's not known in general what the "fastest" radius-3 rule is. The two rules above have the feature that they "do their job" in a fairly "simple-looking" way. But there are also rules like the following that do their job in a "more ornate" way:

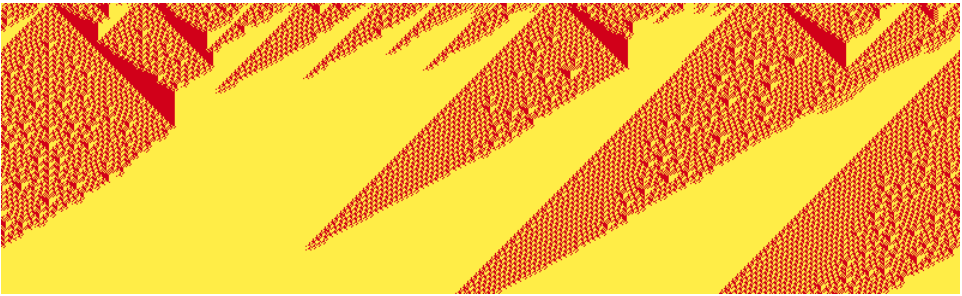

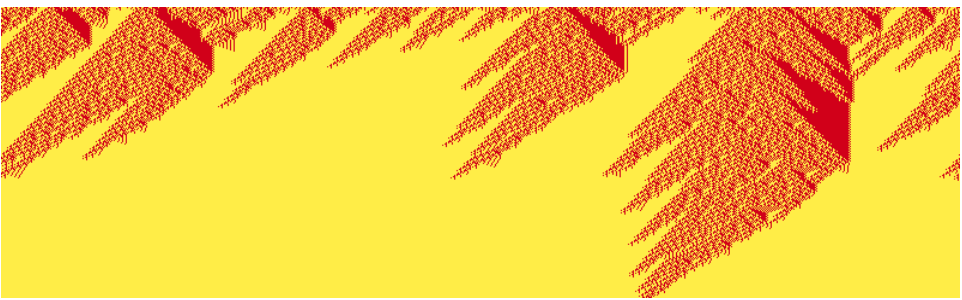

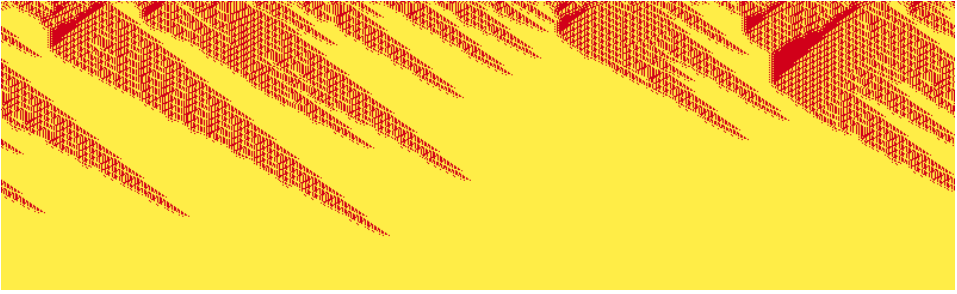

Human-engineered rules (like the first one above) almost inevitably work in simpler and more "understandable" ways. But experience elsewhere (such as with optimal sorting networks) suggests that optimal rules will often be ones that don't look simple in their behavior, and that can't realistically be constructed by standard engineering methods, and essentially just have to be found "experimentally" by searching the computational universe of possible rules.

A notable feature of particularly the earlier rules we looked at is that they show a small number of types of very distinct "domains" with definite walls or boundaries between them. And in many ways such walls can be thought of as being like localized structures, "defects" or "particles". But for our purposes here what tends to be important is whether these particles move around, and whether they annihilate each other to leave a uniform "consensus" final state.

In the simple majority rule it's inevitable that there are static domain walls:

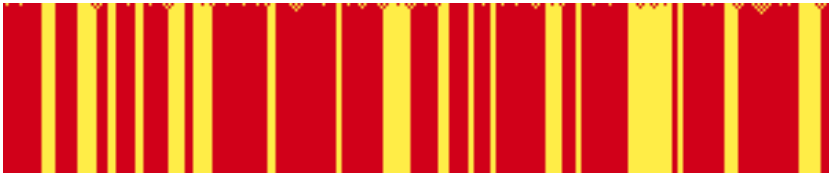

The reason is that as soon as a domain is larger than the cellular automaton neighborhood, a cell at the boundary of the domain will inevitably see a balanced number of cells of each color on the two sides of the boundary. So the cell itself will act as a "tie breaker", and will always decide to stay its own color—thereby making the domain boundary stay as it is.

So what if we have a longer-range rule, that samples more distant cells? With range-2 (i.e. a 5-cell neighborhood) "wrong domains" with widths below 4 disappear:

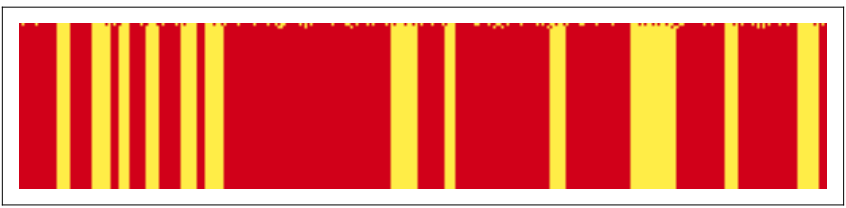

Things work a bit better if the cells being sampled aren't adjacent, but are for example in the pattern ▫-▫▫-▫-▫:

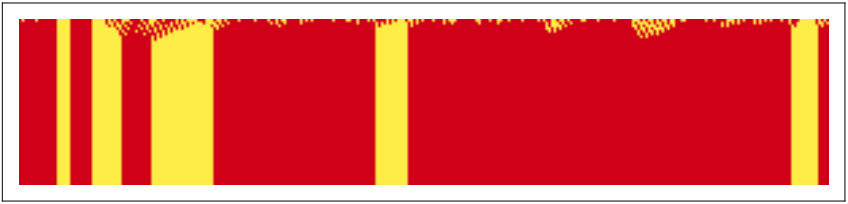

But no finite-size sampling with the pure majority rule will remove all domains. What about the GKL rule? This rule actually only samples 5 cells, but its "extremities" are at distance 3. So can we "improve" it by having it sample more distant cells?

Here's a comparison of a few cases (the first one is the original):

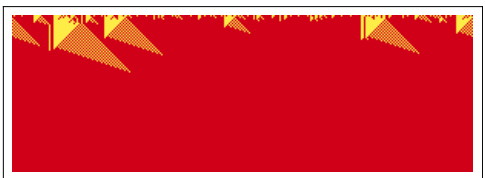
{-3,-1,0,1,3}

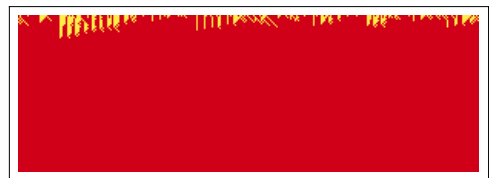
{-5,-1,0,1,5}

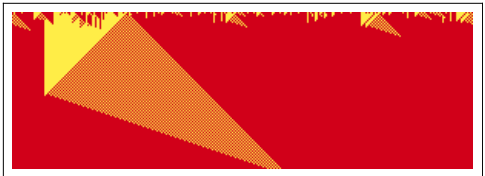
{-3,-1,0,1,5}

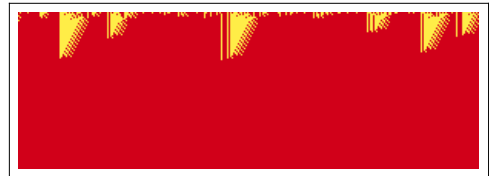
{-3,-1,0,2,4}

Here we've only discussed cellular automata with two possible colors for each cell. We could also consider rules that involve other "helper" colors that either disappear before the final state is reached, or define additional consensus states.

## But Does It Always Work?

We've seen that there are 1D cellular automata that—at least in the examples we've looked at—achieve "majority consensus". But given a particular rule, will it always reach consensus, or are there exceptions?

As a first way to get a well-defined version of that question, we can consider finite cellular automata, say with a total of $n$ cells, and cyclic boundary conditions. There are a total of $2^n$ possible configurations in this case, and we can represent all possible paths of evolution of the cellular automaton using a state transition graph.

Here's the graph for the GKL rule we discussed above, for the case $n = 5$. Each node in the graph is colored according to whether its "red-cell fraction" is above or below $\frac{1}{2}$. And what we see in this case is "perfect density classification" or "perfect consensus", with all states correctly leading to all-red or all-yellow states:

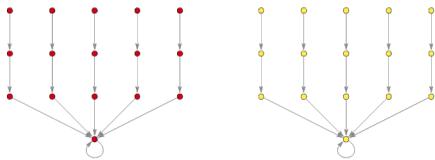

But as soon as we look, for example, at $n = 7$, we immediately see a problem:

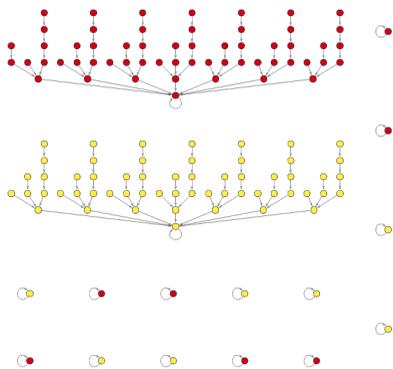

The states

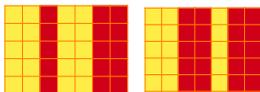

and their cyclic variations "get stuck" and do not achieve consensus. At size 11 there's another issue: now a few states that should have achieved "consensus 1" actually go to "consensus 0":

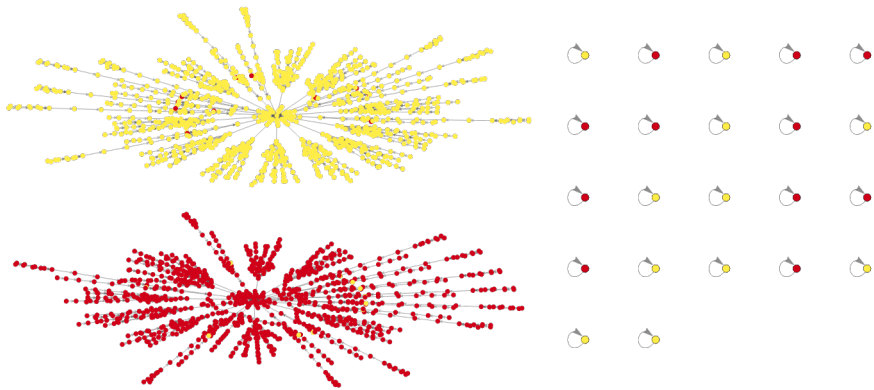

The states that "get to the wrong consensus" here all turn out to be cyclic variations of the following

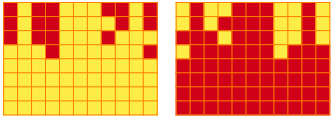

where in the first case there are 6 ■ cells and 5 ▢, yet the final state is all ▢, and in the second case it's the other way around.

And it turns out that there's actually a general problem: one can prove that there's no rule that can perfectly achieve "majority consensus" on a finite array with cyclic boundary conditions.

What about on an infinite array? Here it's possible to achieve "perfect majority consensus" for all but a set of "special initial conditions" with measure 0. An example of such a "special initial condition" is an infinite repetition of either of the two blocks shown above. These initial conditions—instead of going to consensus—will just remain fixed with time.

If initial conditions are generated "at random", with the value of each cell being chosen according to certain fixed probabilities, then there's effectively zero probability of getting one of the "exception" initial conditions. And even though the "tapers" might be arbitrarily long, there's no chance of not eventually reaching a consensus state.

But this conclusion depends on the idea that initial conditions are really generated "at random". If, for example, they were generated by a definite program, then though the initial conditions might seem "statistically random" with respect to certain tests, it doesn't mean that they won't give special weight to the "exceptional" initial conditions.

## Beyond One Dimension

In one dimension one can explain the fact that certain configurations "get stuck" and don't achieve consensus by saying that in 1D effects can't "get around each other". But in 2D there is no such constraint.

So then what about the "pure 2D majority rule" (totalistic code 56):

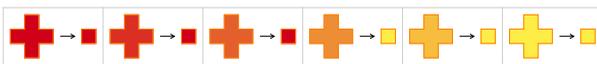

Starting say from 30% 1s we again see that things get stuck:

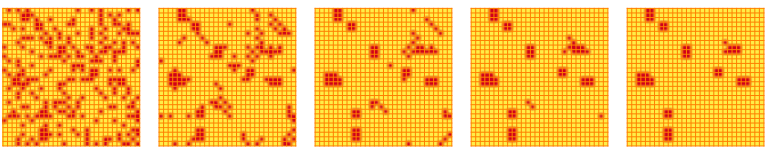

Here is the corresponding evolution shown in 3D, with time going down:

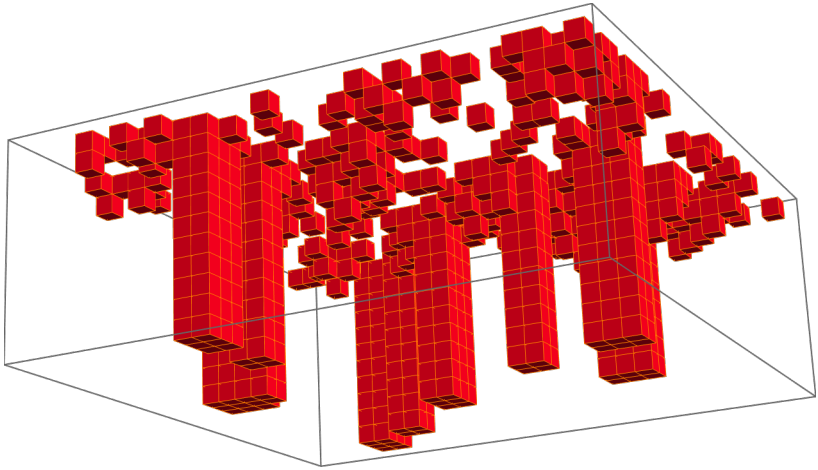

But here is another rule (9-neighbor totalistic code 976):

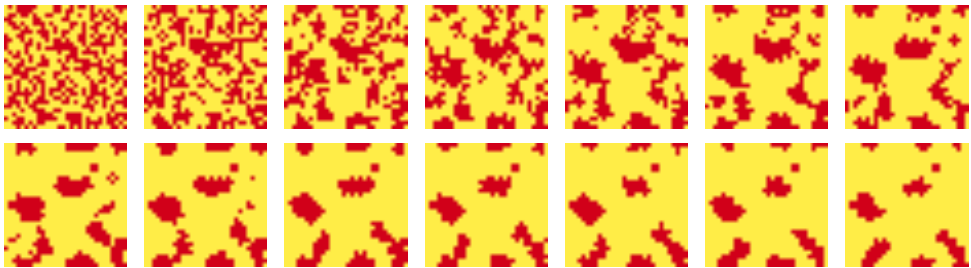

And now what we see is that in this case blob-like domains of the "minority color" get left over, but gradually get smaller. We can see the phenomenon in 3D:

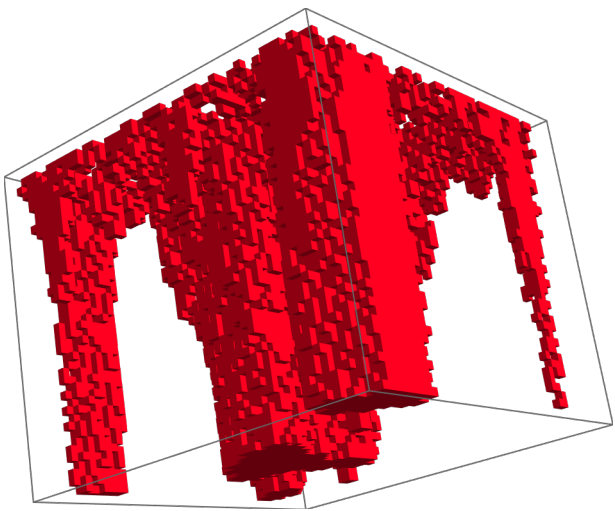

Looking at a spacetime slice in the center, and letting more distant cells "recede into the fog", we see what looks like "diffusive" behavior, with domain walls in effect executing random walks that eventually annihilate:

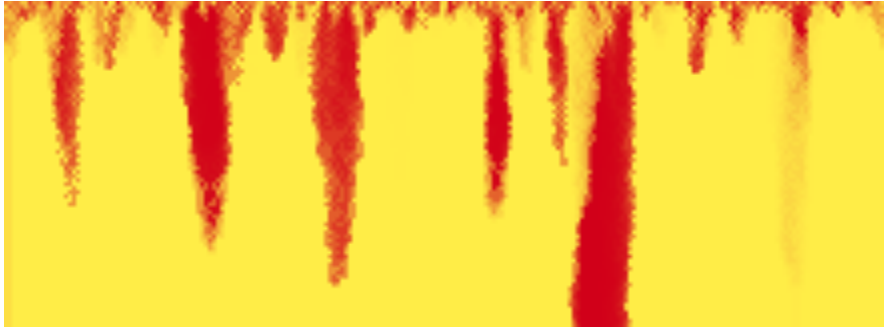

The rule we just saw is close to the majority rule on a 9-cell 3×3 region, except for totals 4 and 5, which are taken to give 1 and 0 rather than 0 and 1. If we use the pure majority rule on the 3×3 region it gets stuck:

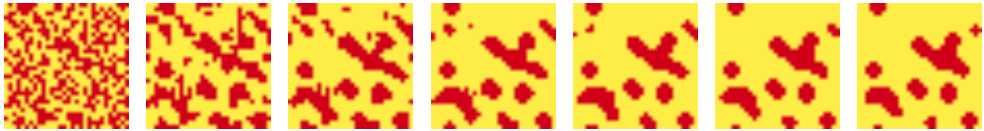

But it turns out to be straightforward to find 2D majority rules that do not get stuck. In fact, basically any majority rule that samples cells in an asymmetric way will work.

As an example, consider a rule that samples the following cells in each 3×3 neighborhood:

Here is the 3D evolution of this rule starting from 45% 1s:

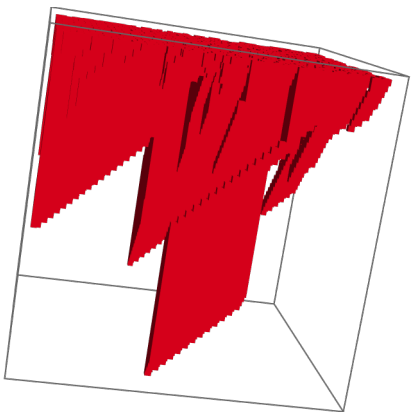

And here is what a spacetime slice looks like:

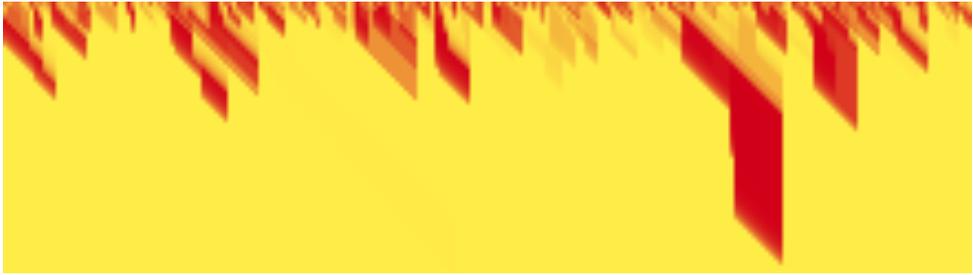

The behavior as a function of the initial density shows a clear transition at 50%:

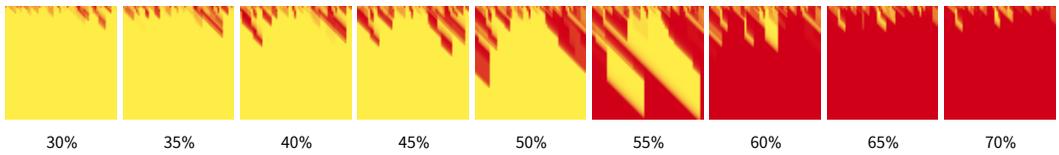

Here are results for different samplings of cells in the 3×3 neighborhood; all successfully achieve consensus:

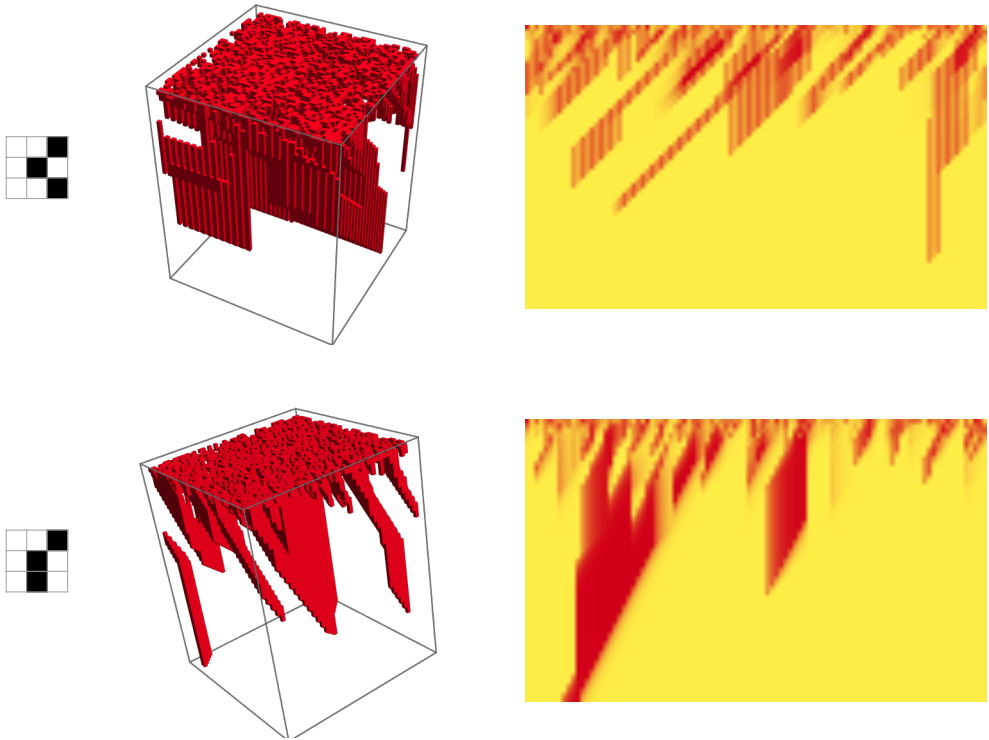

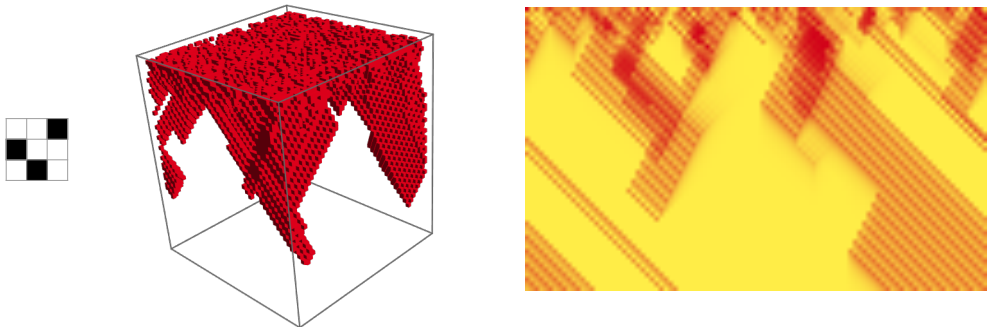

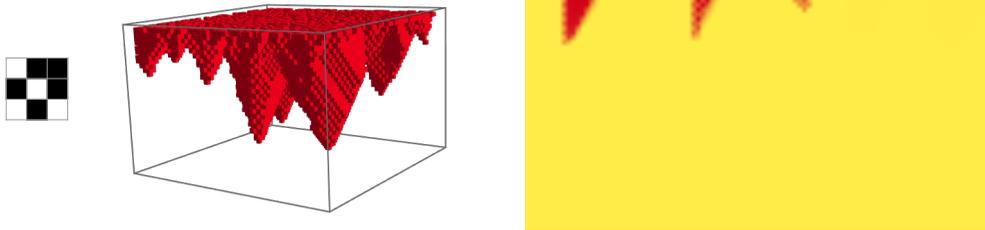

With our original 5-cell "symmetrical" neighborhood we can get very similar behavior by setting things up like the 1D GKL rule:

{{{_, a_, _}, {b_, c_, d_}, {_, e_, _}} :→ If[If[c == 0, a + b, d + e] + c ≥ 2, 1, 0]}

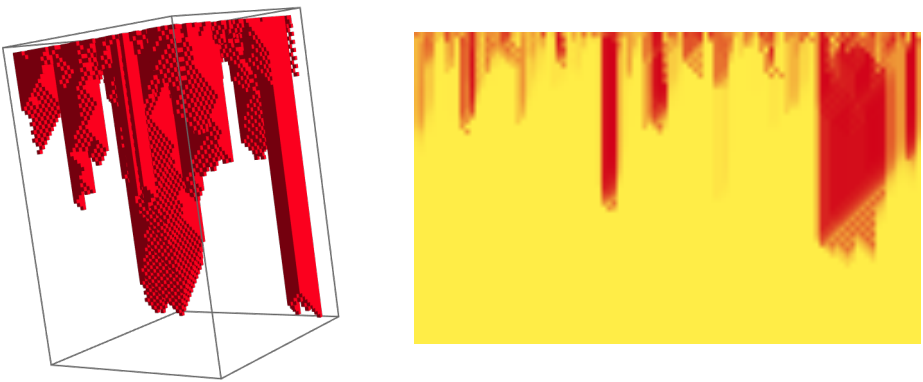

## Cellular Automata with Noise

So far we've assumed that once it's started, the evolution of the cellular automaton is entirely deterministic. But what if there's some "noise" in the evolution—say if the values of cells are randomly flipped with some probability? Here's what happens with the simple majority rule in 1D in this case:

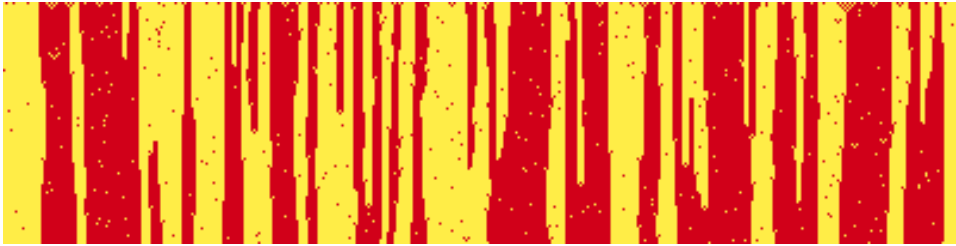

What about the GKL rule? At low levels of noise the rule will typically "fight it off" and still achieve consensus:

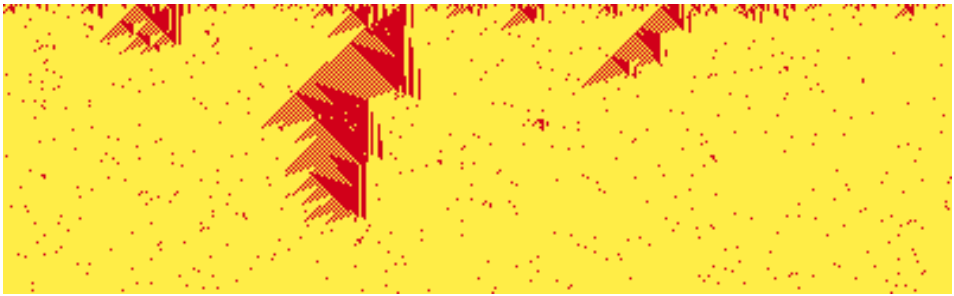

But eventually the level of noise becomes too great, and consensus is typically lost:

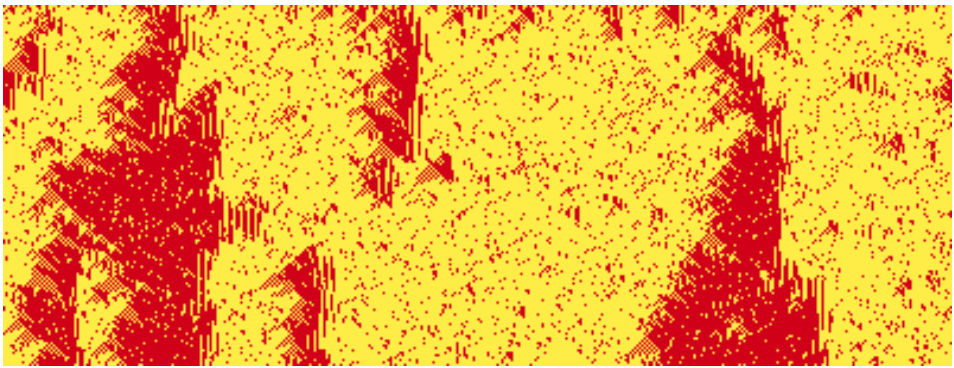

In general, the presence of "noise" turns our system from an ordinary cellular automaton into a probabilistic cellular automaton. (And this, in turn, is equivalent to what's sometimes called directed percolation, or to a spin system that's taken to evolve in time with random updates according to rules with certain weightings. It's also related to what's sometimes been called an "interacting particle system"—in which for example boundaries of regions follow something like an array of random walks, that annihilate when they meet.)

Let's talk in a bit more detail about the overall behavior of the GKL rule. When there's no noise, it shows a sharp transition from final state 0 to final state 1 when the initial density goes from below 0.5 to above 0.5. But what happens when we add noise? We can summarize the result by a classic physics-style phase diagram:

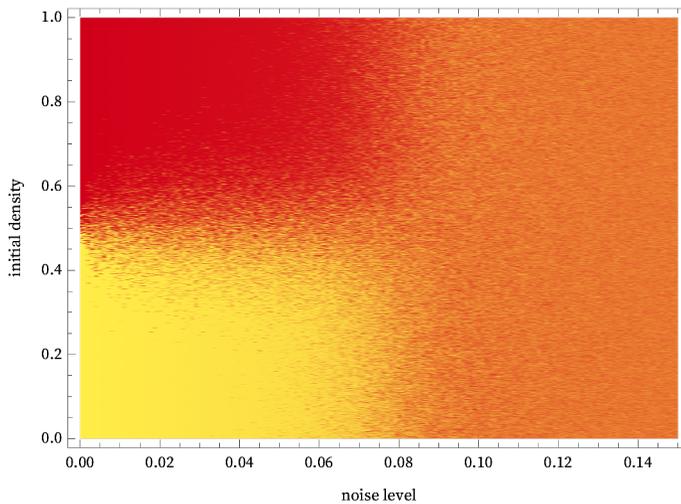

This diagram shows the final density produced by the rule as a function of the initial density and the noise level. At zero noise level, there's a fairly sharp transition as a function of initial density. (It's not perfectly sharp because this diagram was generated by sampling a finite number of initial conditions in a finite region.) And as the noise level increases, the sharp transition seems to survive for a while—until eventually a critical noise level is reached at which it disappears.

Is there a rigorous way to analyze what's going on? Well, not yet. And in fact for a long time it was thought that in the presence of noise any 1D system like this would necessarily be ergodic, in the sense that it would eventually visit all possible states, and certainly not evolve from different initial densities to different final states.

But in the 1980s a complicated cellular automaton was constructed that it was possible to prove would not show such behavior. The system was put together for the purpose of "doing reliable computation even in the presence of noise" and was set up using rather elaborate software-engineering-like methodology. But ultimately it was just a 1D cellular automaton, albeit with an astronomically complicated rule. And the crucial point was that up to some nonzero level of noise, the system could reliably perform a computation—such as achieving majority consensus.

But does one really need a system with such complicated underlying rules to do this? Undoubtedly not. And the situation reminds me of what happened with the problem of ordinary computation universality in cellular automata. Back in the 1950s it seemed one could achieve this with a very complicated setup, constructed in an engineering-like way. But now we know that actually even one of the very simplest conceivable 1D cellular automata—rule 110—is already computation universal. And in fact the Principle of Computational Equivalence implies that whenever we see behavior that is not obviously simple, we can expect computation universality.

Of course, it doesn't seem like we should need to have computation universality to get distributed consensus—though the Principle of Computational Equivalence suggests that computation universality is "cheap" so might in effect "come along for free" with rules that have other necessary properties. (And by the way, this isn't a trivial issue, because when systems are capable of universal computation there's the potential for them to "do something one can't predict", including, for example, break out of some computer security constraint one thinks one's defined.)

But knowing that there's a very complicated cellular automaton that achieves distributed consensus even in the presence of noise makes one wonder what the simplest cellular automaton which can do this might be. And based on my previous experience, I would expect it'll be very simple—like the GKL rule—though it may be very difficult to prove this.

It might be useful to make a few remarks about the whole issue of "noise". In a sense when one says there's "noise" in a system one's saying that the system is "open", and there's something coming from "outside" it that one can't predict. But as an "approximation" one can imagine just having some pseudorandom generator of noise—like the rule 30 cellular automaton. And then one once again has a "closed" system, to which one can immediately apply thinking based for example on the Principle of Computational Equivalence.

But what about "truly unpredictable noise"? To say that this is present is to say that there are different paths of history that the system could follow, and one doesn't know which one will be followed in any particular case. Informed by our Physics Project, we can represent these possibilities by defining a multiway graph, in which there's a branch whenever two different states are generated, depending on the noise.

But in addition to branches, there can also be merges in the multiway graph. And in the rather trivial case of a cellular automaton with the identity rule, allowing every possible individual cell to be flipped by noise, we get the multiway graph:

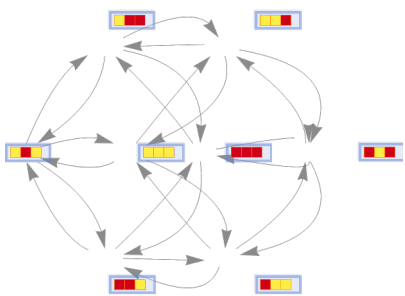

Here's what happens if we apply the majority cellular automata rule (rule 232) after each "noise flip" (and go a total of just 2 steps):

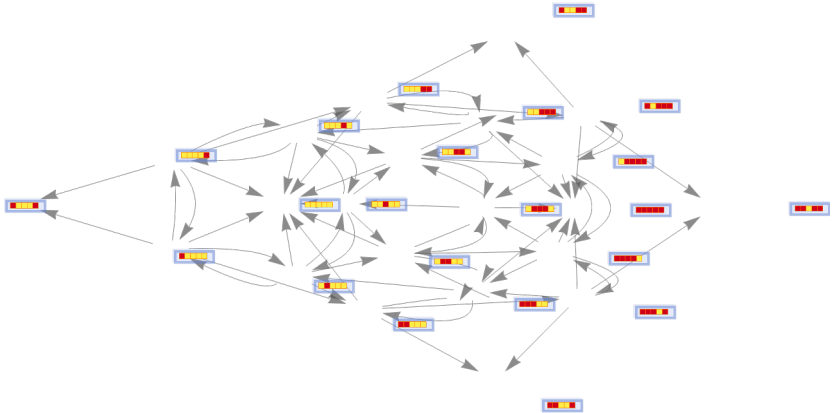

Going more steps, with thicker edges representing more updating events connecting the same states, we get:

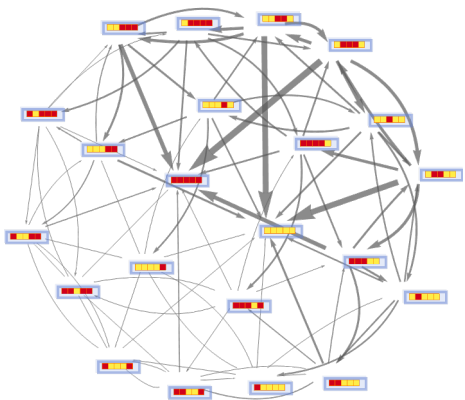

There are several subtle limits to be taken. The size of the cellular automaton is being taken to infinity. The number of steps is also being taken to infinity (though slower). And by saying that there is only a certain "density of noise" we're effectively taking limits on the relative weightings of edges.

To have a system which achieves consensus even in the presence of noise, only particular attractors must survive in these limits. But quite what kind of underlying rule is necessary for this we don't know—though my guess is that it will ultimately be surprisingly simple.

Will computation universality "come along for the ride"? I don't know, but I wouldn't be surprised if it did. Though it's worth understanding that the definition of computation universality in a multiway system like this is somewhat subtle. (I recently discussed it in the context of multiway Turing machines, but there are still more issues when one's interested in probabilities and "probabilistic weightings" of different paths.)

# "Purposeful Attacks" on a Cellular Automaton

We've just talked about the effects of "random noise" on consensus in a cellular automaton. But what about "noise" (or "errors") that are "purposefully introduced"? Is there a pattern of some potentially small number of errors that will, for example, flip the consensus result?

One version of this question—reminiscent of adversarial examples in neural networks—is just to ask what changes need to be made to an initial condition to "flip its result". Or, put another way: let's say one has a system (like the GKL rule) that basically achieves correct consensus for almost all randomly chosen initial conditions. Now we ask the question of whether there is a systematic way to tweak a given randomly chosen initial condition to make it "lead to the wrong answer". (One can think of this as a bit like asking whether one can find a nonce that will make a cryptographic hash come out in a particular way.)

Needless to say, there are many subtleties to this question. What do we mean by "random initial conditions"? Presumably a periodic state wouldn't qualify. What kinds of "tweaks" can we make?

Something that might conceivably happen is that there's a certain behavior with "ordinary" initial conditions, but there's some special "seed" that—if it occurs—will produce unbounded ("tumor-style") growth that eventually takes over the system, as in this simple example from rule 122:

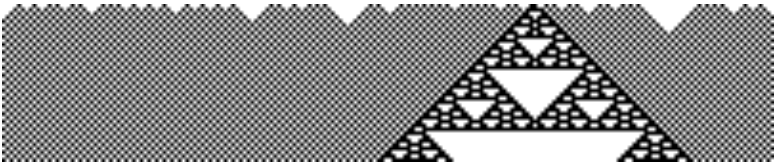

If instead of just "attacking" the initial conditions one allows the possibility of, say, changing the value of a particular cell on every step, it's easy to end up, for example, with "immovable lumps" that in effect prevent "full consensus":

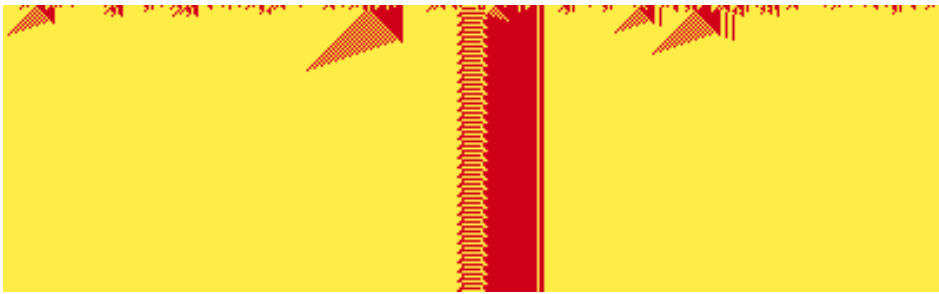

But what if one considers making changes at a small number of places in the ongoing evolution—say in effect adding a little "intelligent noise" at locations carefully computed from the actual pattern of evolution? Will the system always be able to "heal itself" from such "Byzantine tampering", in effect "correcting few-bit errors"? Or is there some particular "vulnerability" that can be exploited to "corrupt" the final results with just a few carefully chosen changes?

One can think of the two consensus final states as being attractors, whose basins of attraction include all initial conditions above or below density $\frac{1}{2}$. Alternatively, one can think of the cellular automaton as "solving the classification problem" of "recognizing the initial density". And perhaps there is some way to extend the cellular automaton to a neural net with continuous weights, and then use machine learning methods to iteratively find minimal places where weights can be changed.

## Graph Cellular Automata

In an ordinary cellular automaton, values are assigned to cells laid out in a definite grid. But as a generalization one can allow the cells to lie at the nodes of a graph—and then to take the neighbors on the graph to define the neighbors to be used in the rule:

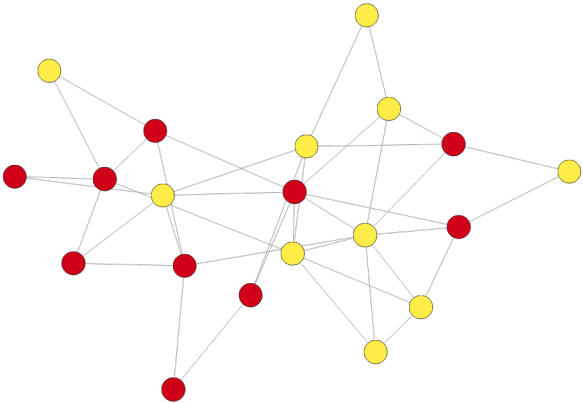

There is one immediate issue here. In the basic definition of a standard cellular automaton, the rule "takes its arguments" in a definite order. But if one's dealing with an ordinary graph (as opposed to, for example, an ordered hypergraph), all one knows is what nodes are connected to a given node—with no immediate ordering defined.

And this implies constraints on the type of cellular automaton rule we can use. One can think of setting up a "geodesic ball" around each node in the graph. Successive "shells" contain nodes that are successive graph distances away from a given node. But the cellular automaton rule can't distinguish which "position" a given cell is at within a particular shell; all it can do is count the total number of cells in each shell that have a given value.

If the graph is vertex transitive, so that the graph structure around every node in the graph is the same (as for a Cayley graph), then the cellular automaton rule can basically contain a fixed table of results that depend only on the number of cells of each value in each shell. But for a general graph the rule for the cellular automaton must allow for arbitrary numbers of cells in each shell.

And one case where this happens to be straightforward to do is for the simple majority rule. So here's an example of this rule applied to "geodesic shells of radius 1" in the graph above:

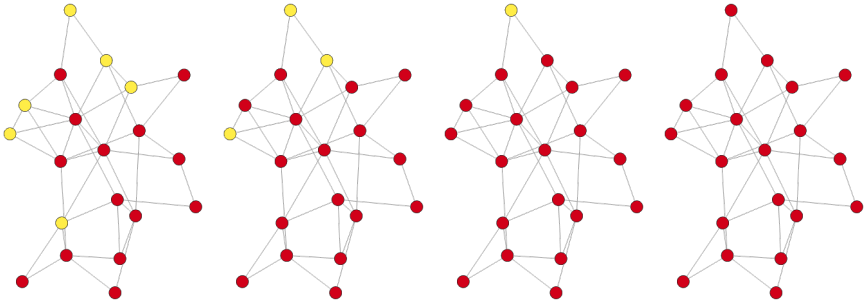

So what globally happens with this rule? For the graph

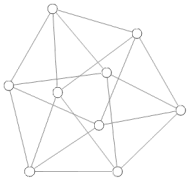

here's what the state transition graph looks like when the rule is applied, whereas above nodes are colored according to which value is in the majority:

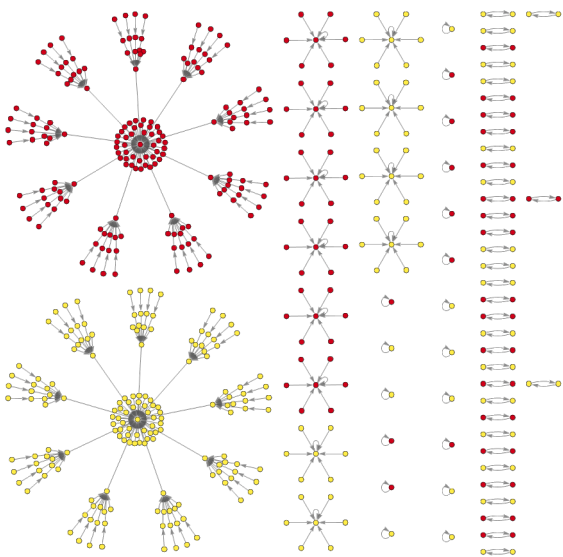

No doubt there are general results that can be proved about the "success rate" for the majority cellular automaton rule on graphs. But experiments tend to suggest that the rule does much better on graphs than it does on regular arrays.

Presumably there are graph-theoretic features of the underlying graph that affect the performance. Higher connectivity presumably helps, not least because it tends to avoid "bridges" where colors can be balanced on "all sides" of a particular node. Lack of symmetry also probably tends to inhibit the appearance of cycles. And in general one can think of the "spreading of consensus" as being at least somewhat like a percolation process.

For ordinary cellular automata, it's clear what it means to ask about the "infinite-size limit". But for graphs it's only immediately clear when one's dealing with some readily extensible family of graphs (like grids or torus graphs or various Cayley graphs). And for arbitrary "random graphs" the results will probably depend significantly on the graph distribution used.

In our Physics Project we have been concerned with large graphs that can be "grown" according to local rules. We expect such graphs often to show certain "statistical regularities" in the "continuum limit". In our project, we characterize the structure of these graphs by looking for example at the growth rates of volumes of geodesic balls, and identifying things like dimension and curvature from them. So what will happen if we run a majority rule cellular automaton on a large graph that has certain "geometrical" properties?

Essentially we need to ask what the "continuum limit" of the majority rule cellular automaton is. The grids used in ordinary cellular automata are too special for them to achieve any kind of generic such limit. But on "geometrizable" graphs, it's more reasonable to expect such a continuum limit.

We can try considering a 1D example. The initial values are then just given by a continuous function of position:

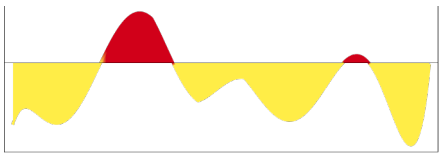

The "consensus result" in this case should be a constant function whose value is effectively the sign of the integral of this function. But what kind of integro-differential-algebraic equation can reproduce the time evolution isn't clear.

Going back to majority cellular automata on graphs, it's worth noting that if the edges of the graph can be assigned both positive and negative weights, then the system is effectively like a synchronous version of a neural net. The analog of this on a regular grid (which is structurally like a spin glass) is then known to show various features of computational irreducibility.

Instead of thinking about underlying graphs with weighted edges, we can consider cellular automaton rules that don't just involve pure nearest-neighbor majority. For example, we could consider rules that have different weights for geodesic shells of different radii (much like the activation-inhibition cellular automata used to model things like biological pigmentation patterns).

But is it really true that only totalistic rules based on geodesic shells can be used for graph cellular automata? To do more than this requires in effect defining "directions" in the graph. But our Physics Project has provided a variety of mechanisms for doing just this—and this in principle for setting up non-totalistic graph cellular automata.

## Asynchronous Updating

An important feature of cellular automata is the assumption that all cell values are updated "simultaneously" or "synchronously" in a definite series of steps. But in practical examples of distributed consensus one's often dealing with values that are instead updated asynchronously. In effect, what one wants is to "break down" the synchronous updating of an ordinary cellular automaton into a sequence of updates of individual cells, with the order of these updates not being specified by any particular rule.

So an obvious first question is: "Does it actually matter in what order these individual updates are done?" And sometimes it doesn't. Here's an example. Instead of an ordinary cellular automaton, consider a block cellular automaton in which at each step pairs of values adjacent cells are replaced by new values:

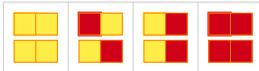

For synchronous updating, we might apply these rules in a systematic "brick-like" pattern. But to study asynchronous updating, let's just apply these rules in random positions at each step. Here are a few examples of what can happen:

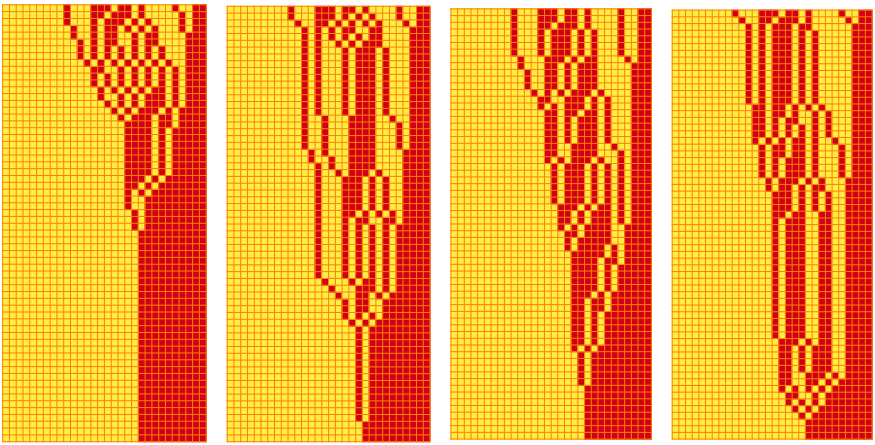

And the notable feature is that even though the specific evolution in each case is different, the final result is always the same—in this case just corresponding to having all 🟨 sorted before 🟥.

It doesn't work this way for all rules, but for this rule, regardless of the intermediate states that are produced, there is always eventual consistency in the final result.

As it turns out, this kind of phenomenon is crucial in our Physics Project. And indeed the generalization that we call "causal invariance" is what leads, for example, to relativistic invariance. But from the formalism of the Physics Project we also get a general approach to asynchronous evolution: trace all possible "update histories" using a multiway graph. Here is the multiway graph for the simple sorting rule above:

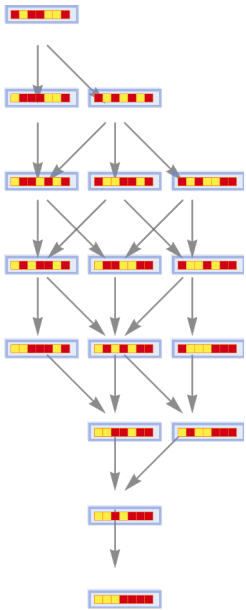

As expected, all possible update histories eventually converge to the same final state.

So what about the majority rule cellular automaton? It doesn't always show eventual consistency, as this example shows:

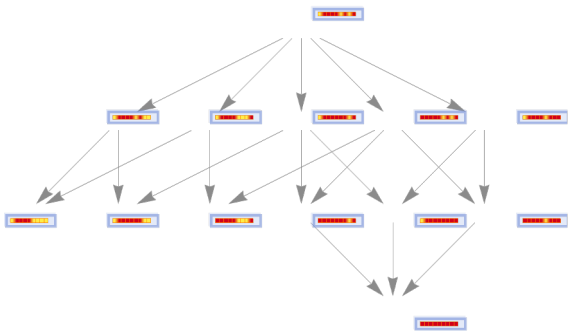

So this means that in general it matters in what order asynchronous updates are done, or, in effect, what "path of history" is taken. But to get a sense of typical behavior, we can consider random sequences of updates. Here's an example of what one gets, doing one update per step:

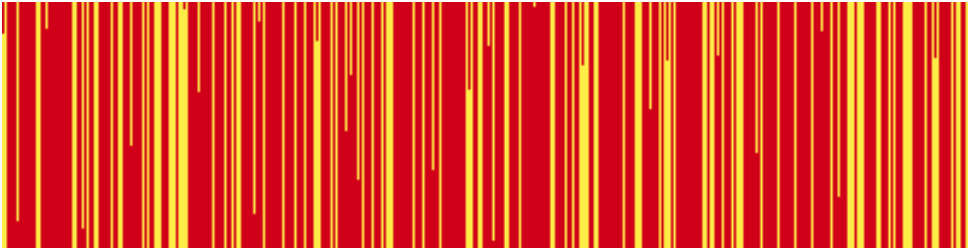

And here's the corresponding result if we do 20 updates per step:

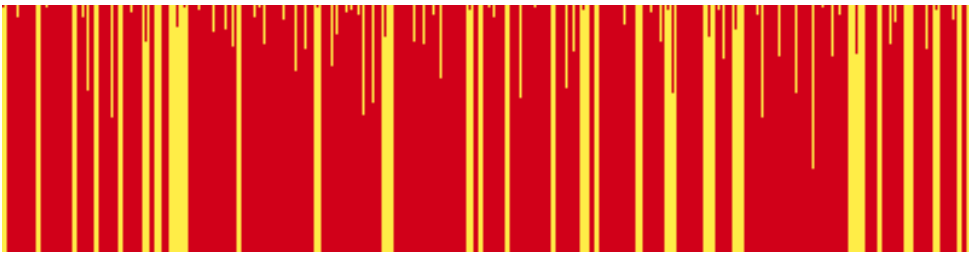

We might have thought that asynchronous updating would add enough randomness to "break ties" and prevent things getting stuck. But in fact it's not hard to see that the results are in this case in the end no different from synchronous updating.

What about for something like the GKL rule? Here are asynchronous results for it, now with 50 updates per step (with initially 60% ■):

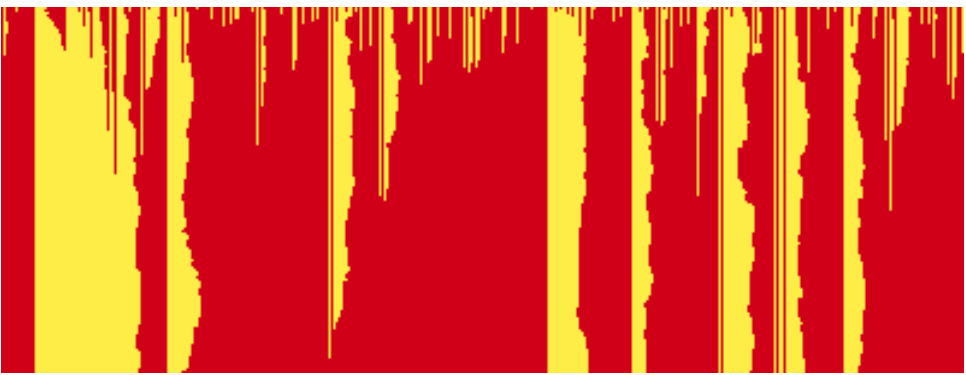

And unlike what we saw above for synchronous updating when we added noise, the change to asynchronous updating seems to completely destroy "majority consensus" for this rule. Note that using more updatings per step does not improve the result (which should in the end be ■ not ▢):

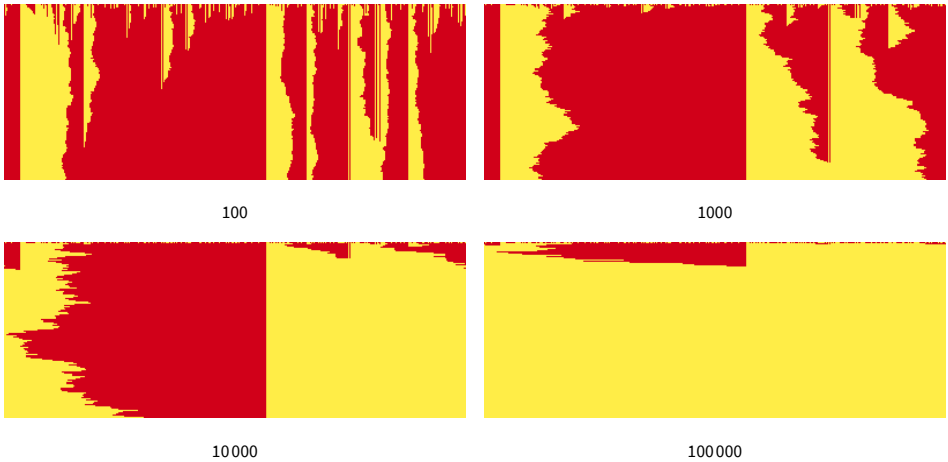

So what happens if we search for rules that achieve consensus asynchronously? In the nearest-neighbor case, the simple majority rule does best, although it's basically no good.

Here are results for a few rules found by searching a million range-2 rules:

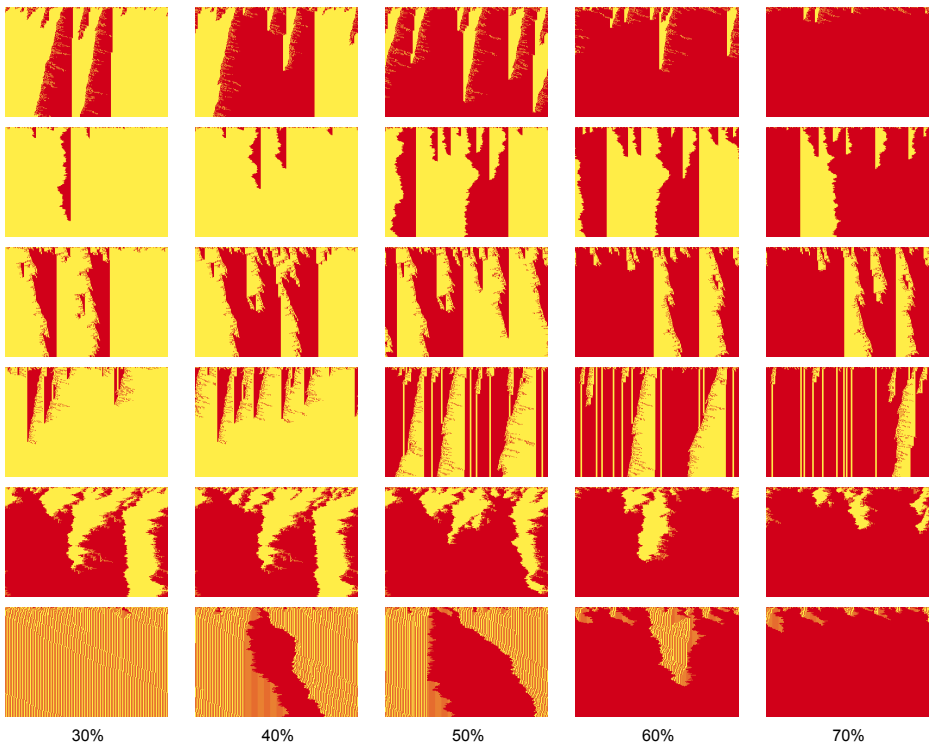

These rules were the top performers in terms of having "closest-to-majority-consensus" average behavior. In the pictures here, an average of 2 updates per cell is being done between successive rows.

If we plot the final density as shown in these pictures against initial density, here are the results for the first 3 rules (with rule numbers 4272826020, 4242057736, and 4265795970):

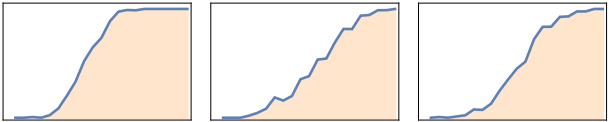

In a perfect consensus rule, these would be step functions at $\frac{1}{2}$—and one can expect that these results may get closer to that with larger numbers of cells and steps.

In an ordinary, synchronous cellular automaton, every cell is in effect updated at every step, and the graph of "causal relationships" between "updating events" forms a trivial grid. But in an asynchronous cellular automaton the graph is sparser—with a particular updating event being causally connected to the previous event that happened to update that cell.

But with the setup we have so far, this causal graph depends only on which cells are updated, not on what their colors might be. And with random updates, the causal graph will basically be like a "random meshing" of the spacetime structure of a system—so that for example for a cellular automaton with cyclic boundaries it becomes an approximation to a tube:

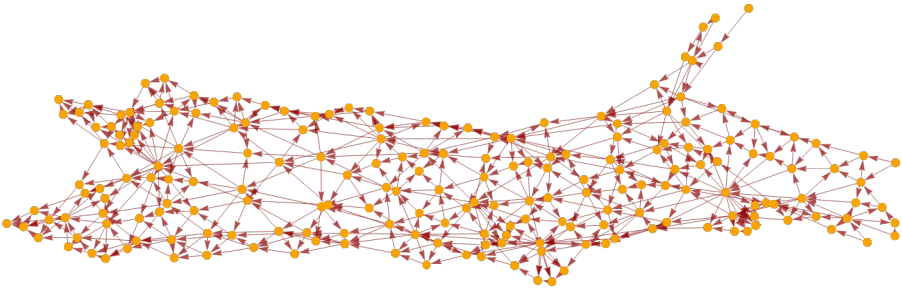

Note that this is just a causal graph for a "single thread of history", associated with a particular sequence of updating events. We can also imagine constructing a multiway causal graph that records the causal relationships both within and between different possible threads of history.

# Dynamic Connectivity

Just as we can consider asynchronous updates in ordinary cellular automata, we can also consider them for graph cellular automata. But once we're considering asynchronous updates on graphs, we can go still further, and consider not just updating "values at nodes" of a graph, but also the graph itself. And in this case, we're basically dealing with the so-called Wolfram models of our Physics Project.

As a kind of bridge to such models, let's consider using them to represent a majority graph cellular automaton. We imagine setting up a hypergraph where all that exists is connectivity of the hypergraph, so "values" in the cellular automata have to be represented by connectivity structures—say with a 0 corresponding to a unary hyperedge, and a 1 corresponding to a ternary one (binary hyperedges are used to make "spatial" connections in the hypergraph).

With this setup, the majority rule becomes a hypergraph transformation rule:

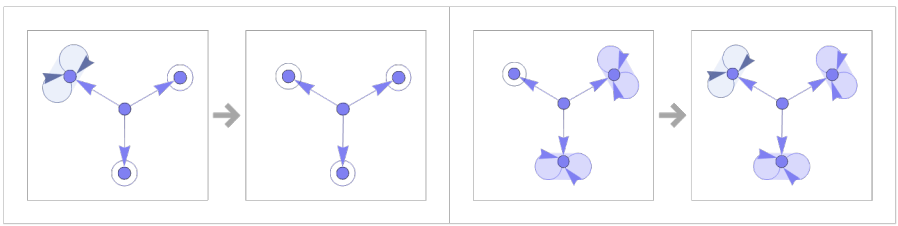

Running this from a particular initial hypergraph, we see consensus achieved in a few steps:

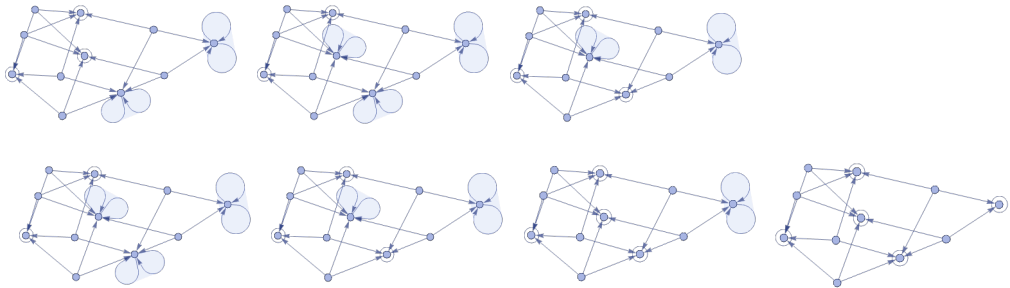

Here is a slightly larger example, that again succeeds in achieving consensus:

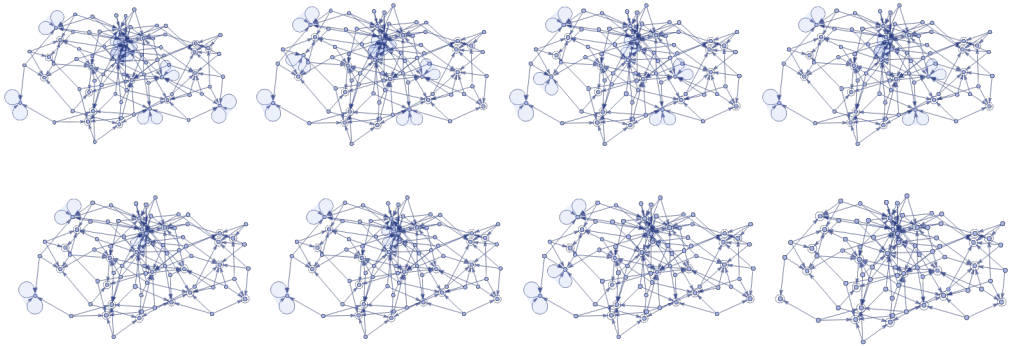

The particular rules we're using here move around the unary and ternary self-loop hyper-edges, but do not affect the "backbone" of the hypergraph. And just as for our earlier examples with ordinary graphs, the simple majority rule doesn't always succeed in achieving consensus.

But now that we have formulated everything in terms of hypergraphs, it's straightforward to have rules that not only change "colors" but also change the underlying structure. As a very simple example, consider adding a "structural rearrangement" case to our rule:

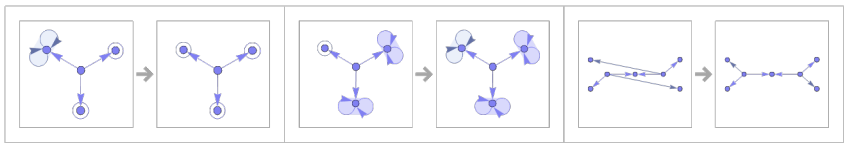

Now in addition to moving around "colors", the rule continually restructures the whole hypergraph:

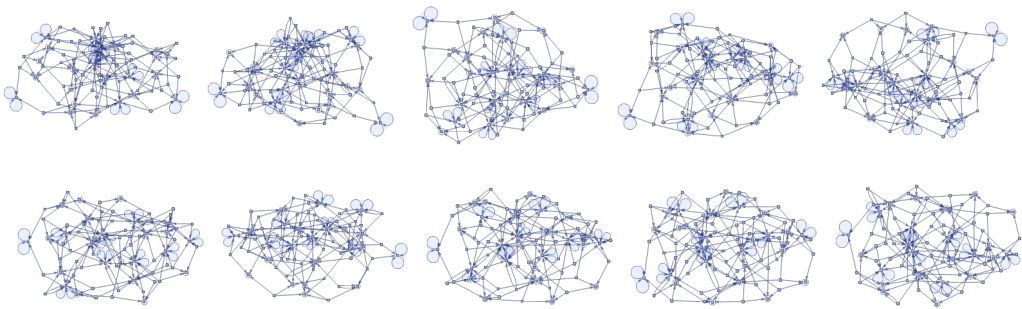

Ultimately this is something very close to our Physics Project. We can imagine encoding values in certain localized structures in our hypergraph—just as we imagine that particles (like photons or quarks) in physics correspond to something like "topological obstructions" in the hypergraph that represents physical space. And in these terms one can imagine formulating questions about consensus in terms of some kind of generalization of conservation laws for particles.

# What's Left to Figure Out

The problem of distributed consensus is in many ways a tantalizing one. The most obvious approach to it—with the simple majority rule—gets a fair distance, but has definite limitations. And as we've seen here, in specific, well-controlled situations there are much better rules—and setups—that can be used. But we don't yet know robust, general, efficient solutions.

One might imagine that to find one would just take "inventing the right algorithm" or "writing the right program". But I think it's unlikely that this kind of traditional "engineering" approach will bear fruit. Instead, I think the most promising path forward is to try to "mine the computational universe" for appropriate rules, in the style suggested by *A New Kind of Science*. And I expect that the best rules will be ones that don't have "readily human understandable" behavior, but instead "do their job" in surprising and perhaps elaborate ways that we would never anticipate.

How can we search for these rules? The most important challenge is to have a good definition of our objective with them. There'll always be tradeoffs. How important is an occasional failure of consensus? How important are different features of the distribution of times to reach consensus? How much do we care about the complexity of the rules? And so on.

So given an objective, what's the best way to actually conduct the search? My consistent experience in mining the computational universe has been that the best results come from the most straightforward strategies. More elaborate strategies tend to make implicit assumptions, that prevent the discovery of truly surprising or unexpected results.

A good start is just to do an exhaustive search. It's important to be very careful in pruning it, lest one miss the "unexpected way" that a system can achieve some particular objective. Is it likely to be possible to "incrementally improve" rules, say with genetic algorithms? I'm not especially hopeful. Because to make serious use of what the computational universe has to offer, our rules are likely to need to show computational irreducibility—and this makes it essentially inevitable that the "landscape" of "nearby" rules will be irreducibly "rough", making any computationally bounded incremental improvement unlikely to be successful.

Could we perhaps train a machine learning system to suggest useful rules? It may be possible to do some pruning of candidate rules this way, although inevitably there is some risk of missing the "unexpected rule". And in general the presence of computational irreducibility makes it implausible that an incrementally trained machine learning system will be extremely successful.

One might have thought that something like exhaustive search could never find useful results, because the space of rules is in some sense just too big. But a key discovery from my explorations of the computational universe is that in fact there are surprisingly simple rules that can show rich and sophisticated behavior. And this makes it plausible that one could discover a good solution to the problem of distributed consensus just by appropriately searching the computational universe—and "mining" some rule that can then be used quite generally as a basis for all sorts of practical distributed consensus.

## Some Historical Background

Investigations of what amounts to distributed consensus have a fairly long, if seemingly scattered history. As soon as even somewhat complex electromechanical and electronic systems were being built, the question arose of how to make the whole system behave in a reliable way even if some of its components were unreliable. The simplest answer was to have redundancy, and somehow to "take a vote", and go with the "majority" decision. In the earliest computers (and later particularly in aerospace systems) such a vote was typically between copies of more-or-less complete systems.

But by the beginning of the 1950s there was increasing interest in moving the voting down to the level of smaller components. And in 1952 John von Neumann, in his "Probabilistic Logics and the Synthesis of Reliable Organisms from Unreliable Components", began to give a mathematical structure for analyzing this. Central to his discussion was what he called the "majority organ", which is essentially a component for computing the Boolean majority function:

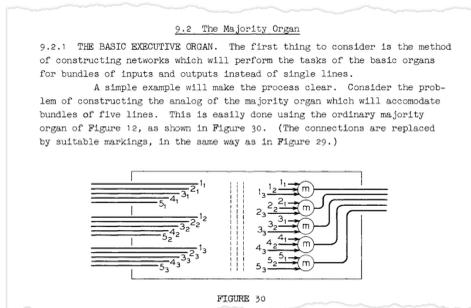

von Neumann imagined building up everything (including the majority organ) from what he called "Sheffer organs"—or what we would now call Nand gates. And as an example of the need for redundancy he says "Consider a computing machine with 2500 vacuum tubes, each actuated on average every 5 microseconds. Assume a mean free path of 8 hours between errors is desired." Somewhat mysteriously he then assumes the very high (even for the time) error rate $\epsilon = 0.005$ and concludes that to operate reliably "the system should be multiplexed 14,000 times". (von Neumann goes on to talk about errors in brains, as modeled by neural nets. Rather implausibly he states that in brains "errors are not ordinarily observed", and concludes from this that the multiplexing factor must be about 20,000—which he viewed as being consistent with what was then known about actual brains.)

Fortunately for the history of computing von Neumann's example error rate turned out to be very wide of the mark—and once vacuum tubes were replaced by solid-state devices the problem of component failures in computers more or less disappeared (though it reappears in modern thinking about molecular-scale computing). In communications systems (and, to a lesser extent, storage devices) errors were always still important, and this led by the 1960s to increasing work on error-correcting codes.

But perhaps because the transition to solid-state electronics happened more slowly in the Soviet Union interest in the problem of getting reliable results from unreliable components lasted much longer there. And while in the West, such issues tended to be thought of as matters of applied engineering, in the Soviet Union they were much more considered a matter of pure mathematics. (In the West, there was also in the 1950s the rather amorphous idea of "cybernetics", which was initially considered ideologically inappropriate in the Soviet Union, but was later adopted there, and turned in a much more mathematical direction.)

But in addition to questions coming essentially from the construction of machines (or brains), there was an initially quite separate strand of questions coming from physics. A very basic observation in physics is that materials undergo so-called phase transitions. For example, as one heats up water, there is a definite temperature at which all the molecules "together decide" to make the transition from liquid to gas.

A somewhat more subtle version of the same kind of thing occurs in magnetic materials like iron. Below a certain temperature, electron spins associated with all the atoms in the material tend to line up. But in what direction? Somehow a "consensus direction" is selected—that defines the macroscopic direction of the magnetic field produced by the material. And in the 1920s the Ising model was suggested as a simple model for this.

But before getting to phase transitions, there was a more basic physics question of how microscopic discrete elements like atoms could in general lead to macroscopic phenomena. And a key part of this question had to do with the understanding the motion of molecules and how this could lead to "thermodynamic equilibrium". The whole story of the foundations of "statistical mechanics" got quite muddled (and I think it's only quite recently, with computation-based ideas, that we've finally been able to properly sort it out). But particularly in the first few decades of the 1900s the key idea was thought to be ergodicity: essentially the notion that the equations of motion of molecules will lead them eventually to visit all possible states of a system, thereby, it was argued, making their behavior seem random.

It was difficult to establish ergodicity mathematically. But beginning around the 1930s this was a major emphasis of the field of dynamical systems theory. Meanwhile, there were also difficulties in understanding mathematically how phase transitions could occur. And one point of contact was that when there's a phase transition, ergodicity effectively has to be broken: the spins in a magnet end up in a particular direction, and don't visit all directions.

At the beginning of the 1960s there was a convergence in Moscow of a considerable number of top Soviet mathematicians (notably including Andrei Kolmogorov) who were variously working on statistical mechanics, ergodic theory, dynamical systems—and some of the mathematical sequelae of cybernetics. And one of the pieces of work that emerged was a paper in 1968 by a then-young math-competition-winning mathematician named Andrei Toom.

The (translated) title of the paper is "A Family of Uniform Nets of Formal Neurons". The paper is cast in terms of formal probability theory and the study of Markov chains. But basically it's a construction of what amounts to a probabilistic cellular automaton, and a proof that even though it's probabilistic, certain aspects of its behavior are "non-ergodic" and effectively deterministic. (It's notable that in 1963 Toom had done another construction: of what's now called the Toom—Cook algorithm for fast multiplication of integers with many digits.)

In traditional statistical mechanics (which was somewhat distinct from ergodic theory) the original focus was on studying "equilibrium" systems, in which different possible configurations (say of the Ising model) occurred with particular weightings. But by the 1950s—especially in work at Los Alamos—the idea arose of sequentially "Monte Carlo" sampling these configurations on a computer. And in 1963 Roy Glauber suggested thinking of the actual dynamics of the Ising model in terms of sequential probabilistic updating of spins.

Meanwhile, somewhat separately, there was increasing study—particularly by American mathematicians such as Frank Spitzer—of the probability theory of collections of random walks, often referred to as "interacting particle systems". And one of the main results was that as soon as nonzero probabilities were involved, ergodicity was typically found.

Apparently independent of these developments the Moscow group in 1969 produced a paper entitled "Modeling of Voting with Random Error". It featured a calculation done on an "electronic computing machine" ("ЭВМ" in Russian) of the probabilistic evolution of a majority model (and quite likely the machine used was a base-3 Setun computer developed by the mathematician Sergei Sobolev):

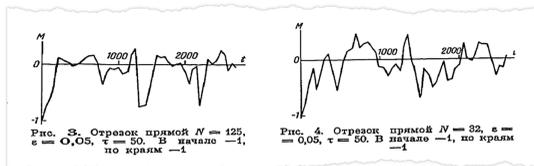

The paper concluded that in 1D, the model was probably always ergodic, but in 2D, for sufficiently small noise level, it might not be.

In 1971 Roland Dobrushin (a student of Kolmogorov's) connected the investigation of ergodicity in these networks with phase transitions in Ising models—which helped define the program of research at his "Laboratory of Multi-component Random Systems" at the "Institute for Problems of Information Transmission" that brought together the Soviet cybernetics tradition (with its work on things like neural nets, Markov chains and formal computability theory) with international work on mathematical physics and ergodic theory.

A typical product of this was the 1976 conference organized by Dobrushin (along with Toom and others) nominally entitled "Locally Interacting Systems and Their Application in Biology"—but actually with very little biology in sight, and steeped in sophisticated mathematics, about things like Markov fields, Gibbs measures and algorithmic unsolvability.

A key question that had emerged was whether a homogeneous array of probabilistic elements (i.e. a probabilistic cellular automaton) could consistently and deterministically store information, or whether inevitably there would be ergodicity that would destroy it.

In 1974 Toom showed that a multidimensional probabilistic cellular automaton could do this—essentially just using a majority rule on a non-symmetric neighborhood to generate a "global consensus state", as we showed above. But the question still remained of whether anything similar was possible in 1D.

Phase transitions in traditional statistical physics don't happen in 1D if microscopic reversibility is assumed—making it seem like it might be impossible to maintain multiple distinct global states. But in 1976 Boris Tsirelson pointed out that at least with a hierarchical arrangement of interactions one could in fact achieve long-range order in a probabilistic 1D system:

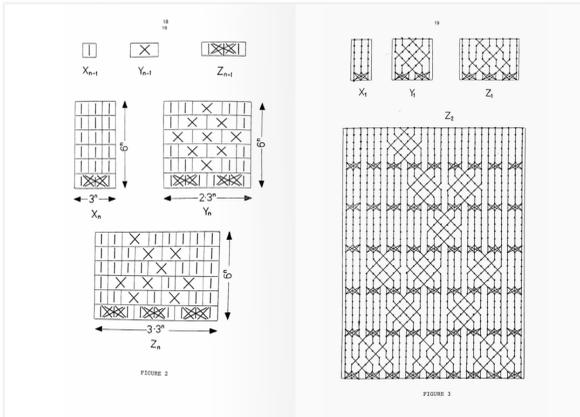

Soon thereafter Georgii Kurdyumov—having at first discussed the undecidability of ergodicity in the 1D case—then argued that there should be a pure cellular automaton that would work.

And in 1978, Peter Gacs, Georgii Kurdyumov and Leonid Levin (all of whom had been in the Kolmogorov orbit) wrote a short paper entitled "One-Dimensional Uniform Arrays that Wash Out Finite Islands" that introduced the "GKL rule" we discussed above. They didn't show any actual pictures of the behavior of the system, but they gave a proof that in the deterministic case the rule leads to two distinct phases, corresponding to the two distinct consensus states. And then they showed the result of a simulation that suggested that even when a certain amount of noise was added the two consensus states would still be reached:

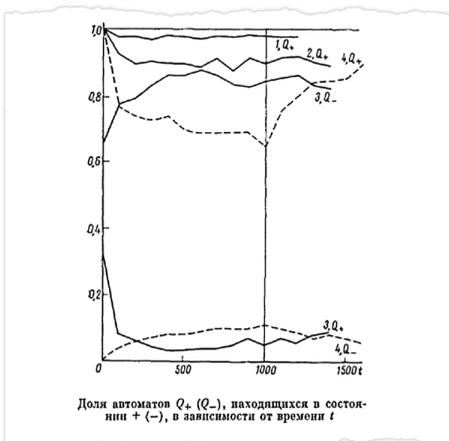

However, what had become known as the "positive probability conjecture" implied that there couldn't in the end actually be non-ergodicity in the 1D case. But in 1983 Peter Gacs came up with what he claimed was a counterexample based on an elaborate construction described in many pages of pseudocode:

It took many years for the proof of this to clarify, with Gacs publishing a final version only in 2001.

Meanwhile, there'd been several other developments. Starting in 1982 my own discoveries about deterministic cellular automata had made 1D cellular automata much more prominent—and had made physicists aware of them. (As it happens, Gacs announced his 1983 result at a conference at Los Alamos I had organized, that I believe was the first ever to be devoted to cellular automata.)

Around the end of the 1980s there was then a burst of activity by several leading mathematical physicists devoted to applying methods from statistical mechanics (and especially from areas like directed percolation theory) to the analysis of probabilistic cellular automata. There was awareness of Toom's work, and for example connections were made between PDEs (like the KPZ equation) and things like the average behavior of "domain walls" in the probabilistic Toom rule.

One can view the process of coming to consensus in a 1D cellular automaton as being like a "density classification" problem: if the initial density of 1s is above $\frac{1}{2}$, classify as 1, otherwise classify as 0. And starting in the 1990s density classification in 1D (deterministic) cellular automata was used as a prime example of a place where algorithms might be discovered by genetic or other search techniques.

In a quite different direction, work on cryptographic protocols in the 1980s had highlighted various models for achieving consensus between agents, even in the presence of adversarial efforts. Meanwhile, there was increasing interest in formal models of parallel computation, their computational complexity, and their fault tolerance. And by the early 2000s there was work being done (notably by Nick Pippenger) on connections between these things and what was known about probabilistic cellular automata, and the possibility of deterministic computation in them.

And this pretty much takes us to the current time—and the new applications of distributed consensus in blockchain-like systems. And here it's interesting to see the rather different intellectual lineages of two different efforts: Yilun Zhang at NKN coming from a statistical physics/computational neuroscience/information theory tradition, and Serguei Popov at Iota coming from probability theory and stochastic processes—as a great-grand-student of Kolmogorov.

## Some Personal Notes

Of all the work I've done on cellular automata and related systems over the past more than forty years rather little has been devoted to the topics I've been discussing here. There are a couple of reasons for this. The most important is that my main interest has been in studying the remarkable richness and complexity that cellular automata and other very simple programs can generate—and in building a paradigm for thinking about this. Yet something like distributed consensus is at some level about getting rid of complexity rather than generating it. It's about taking whatever complicated initial state there may be, and somehow reducing it to a "simple consensus", where there's none of that complexity.

Another point is that at least some of what we've discussed here has concerned probabilistic systems, which I've tended to ignore on the grounds that they obscure the fundamental phenomena of the computational universe. If one didn't know that simple, deterministic rules could do complex things, one might imagine that would have to inject randomness from the outside to make this happen. But the fact is that even very simple, deterministic rules can produce highly complex behavior, that in fact often makes its own apparent randomness.

So that means there's no need to "go outside the system"—and to introduce external randomness or probabilities. And in fact such probabilities tend to have the effect of hiding whatever complexity is intrinsically produced—even if they do "smooth out average behavior" to make things more accessible to traditional mathematical methods.

There are actually some new perspectives on this from our Physics Project. First, the project makes clear the crucial interplay between underlying computational irreducibility, and effectively probabilistic large-scale behavior that can be treated in computationally reducible ways. And second, the project suggests that instead of thinking about probabilities for different behavior, one should think about the whole multiway system of possible behaviors, and its overall properties.

It so happens that when I first became interested in the origins of complexity the first two kinds of models I thought about were spin systems (like the Ising model) and neural nets. But as I tried to simplify things I ended up inventing for myself what I soon found out were one-dimensional cellular automata. Much of my effort was then concentrated in doing experiments on these systems, and in developing theories and principles around the results I found.

But I also tried to do my homework on earlier work. Cellular automata had gone by many names. But leafing through the (then on paper) Science Citation Index I slowly began to piece together some of their history, and soon found things like the paper introducing the GKL rule. In my first long paper on cellular automata (entitled "Statistical Mechanics of Cellular Automata" and published in 1983) I have just a few paragraphs about "probabilistic rules", discussing ergodicity and phase transitions, and referencing the GKL paper.

Over the years I accumulated five thick folders of copies of papers that I labeled as being about "Stochastic Cellular Automata". And I also purchased books. And in writing this piece I was able to just pull off my shelf things like Dobrushin's 1976 book. And in one of those manifestations of the smallness of the scientific world, when I looked in the front of my (apparently used) copy of this book yesterday, what should I see there but the signature of Frank Spitzer—who I had just been writing about!

When I was writing *A New Kind of Science*, both probabilistic cellular automata and what amounts to the problem of consensus did come up, and there are several mentions of such things in the book, notably in connection with my discussion of the "Origins of Discreteness":

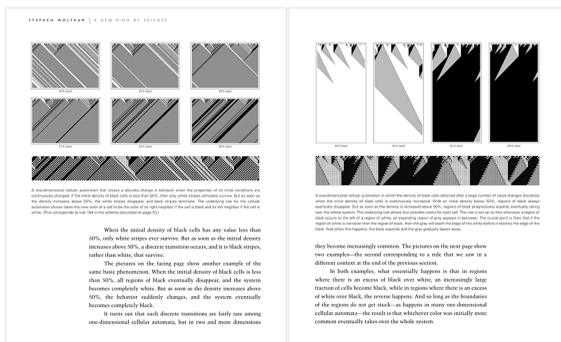

But these things were never a big emphasis of my work, and so it's been interesting here to trace just how the methods I've developed can be applied to them, and to realize that—despite its slightly different presentation—the problem of distributed consensus is in many ways actually a quintessential question that can be addressed by the kind of science that's derived from studying the computational universe.

# A Bibliography of Distributed Consensus with Cellular Automata & Related Systems

44 | Stephen WolframC. Cooper, et al. (2014), "The Power of Two Choices in Distributed Voting," in *Automata, Languages, and Programming,* J. Esparza, et. al. (eds), International Colloquium on Automata, Languages, and Programming 2014, *Lecture Notes in Computer Science* vol. 8573, Springer, 435—446. doi: 10.1007/978-3-662-43951-7_37.

J. Cruise and A. Ganesh (2014), "Probabilistic Consensus via Polling and Majority Rules," *Queueing Systems* 78, 99—120. doi: 10.1007/s11134-014-9397-7.

P. P. B. de Oliveira (2014), "On Density Determination with Cellular Automata: Results, Constructions and Directions", *Journal of Cellular Automata* 9, 357—385.

S. Monica and F. Bergenti (2014), "A Stochastic Model of Self-stabilizing Cellular Automata for Consensus Formation", in *CEUR Workshop Proceedings*, Central Europe Workshop 2014, Sun SITE Central Europe.

C. Cooper, et al. (2015), "Fast Consensus for Voting on General Expander Graphs," in *Distributed Computing*, International Symposium on Distributed Computing 2015, *Lecture Notes in Computer Science* vol. 9363, Springer, 248—262. doi: 10.1007/978-3-662-48653-5_17.

A. Gogolev, et al. (2015), "Distributed Binary Consensus in Networks with Disturbances", *ACM Transactions on Autonomous and Adaptive Systems* 10, 19:1—19:17. doi: 10.1145/2746347.

L. Becchetti, et al. (2016), "Stabilizing Consensus with Many Opinions," in *Proceedings of the Twenty-Seventh Annual ACM-SIAM Symposium on Discrete Algorithms*, Society of Industrial and Applied Mathematics 2016, pp. 620—635. doi: 10.1137/1.9781611974331.ch46.

I. Benjamini, et al. (2016), "Convergence, Unanimity and Disagreement in Majority Dynamics on Unimodular Graphs and Random Graphs", *Stochastic Processes and Their Applications* 126, 2719—2733. doi: 10.1016/j.spa.2016.02.015.

R. Elsässer, et al. (2016), "Rapid Asynchronous Plurality Consensus". arXiv:1602.04667.

D. Griffin (2016), "The Consensus Problem, Cellular Automata, and Self-replicating Structures", Physics Capstone Project, paper 35, Utah State University.

I. Sharma, et al. (2016), "On the Role of Evangelism in Consensus Formation: A Simulation Approach", *Complex Adaptive Systems Modeling* 4, doi: 10.1186/s40294-016-0029-4.

P. T. Tošić (2017), "Phase Transitions in Possible Dynamics of Cellular and Graph Automata Models of Sparsely Interconnected Multi-agent Systems", in *Proceedings of the International Joint Conference on Autonomous Agents and Multiagent Systems, AAMAS*, International Conference on Autonomous Agents and Multiagent Systems 2017, 474—483. doi: 10.5555/3091125.3091195.

P. Gács and I. Törmä (2018), "Stable Multi-level Monotonic Eroders". arXiv:1809.09503.

E. Goles, et al. (2018), "On the Complexity of Two-Dimensional Signed Majority Cellular Automata", *Journal of Computer and System Sciences* 91, 1—32. doi: 10.1016/j.jcss.2017.07.010.

# References

*Links to references are included within the body of this document.*